\documentclass[conference]{IEEEtran}

\def\BibTeX{{\rm B\kern-.05em{\sc i\kern-.025em b}\kern-.08em
    T\kern-.1667em\lower.7ex\hbox{E}\kern-.125emX}}

\usepackage{cite}
\usepackage{amsmath,amssymb,amsfonts}
\usepackage{algorithmic}
\usepackage{graphicx}
\usepackage{textcomp}
\usepackage{xcolor}
\usepackage{makecell}

\usepackage{siunitx}
\usepackage{hyperref}
\usepackage[nolist]{acronym}
\usepackage[flushleft]{threeparttable}
\usepackage{placeins}
\usepackage{svg}                                   
\usepackage[percent]{overpic}
\usepackage[nameinlink, capitalise]{cleveref}

\usepackage{mathtools}
\usepackage{booktabs}
\usepackage[nolist]{acronym}
\usepackage{pbalance} 
\usepackage[export]{adjustbox}

\usepackage{eso-pic}    

\usepackage{orcidlink}

\usepackage[inline]{enumitem}

\usepackage{booktabs}
\usepackage{tabularx}
\usepackage{makecell}
\usepackage{booktabs}
\usepackage{array}
\usepackage{multirow}

\newcommand{\RNum}[1]{\uppercase\expandafter{\romannumeral #1\relax}}   

\DeclareRobustCommand{\IEEEauthorrefmark}[1]{\smash{\textsuperscript{\footnotesize #1}}}

\DeclareSIUnit\db{dB}                           
\DeclareSIUnit\dbi{dBi}                         
\DeclareSIUnit\dbm{dBm}                         
\DeclareSIUnit\watthour{Wh}                     
\DeclareSIUnit\mbps{Mbps}                       
\DeclareSIUnit\kbps{kbps}                       
\DeclareSIUnit\bps{bps}                         
\DeclareSIUnit\msInference{ms/inference}        
\DeclareSIUnit\persquare{\ensuremath{/\square}}

\begin{document}
\bstctlcite{IEEEexample:BSTcontrol} 

\AddToShipoutPictureBG*{
  \AtPageUpperLeft{%
    \put(0,-40){\raisebox{15pt}{\makebox[\paperwidth]{\begin{minipage}{21cm}\centering
      \textcolor{gray}{This article has been accepted for publication in the proceedings of the \\
       International Symposium on Circuits and Systems (ISCAS 2026)\\ 
       } 
     \end{minipage}}}}%
   }
   \AtPageLowerLeft{%
     \raisebox{25pt}{\makebox[\paperwidth]{\begin{minipage}{21cm}\centering
       \textcolor{gray}{ \copyright 2026  Authors and IEEE. 
        This is the author’s version of the work. It is posted here for your personal use. Not for redistribution. \\
        The definitive Version of Record will be published in the proceedings of the International Symposium on Circuits and Systems (ISCAS 2026).
         }
     \end{minipage}}}%
   }
 }


\title{A Miniaturized In-Mouth pH Sensing System for Real-Time Intraoral Telemetry
}

\author{
    \IEEEauthorblockN{
        Lukas Schulthess\;\orcidlink{0000-0002-6027-2927}\,\IEEEauthorrefmark{1},
        Philipp Schilk\;\orcidlink{0000-0002-0487-9513}\,\IEEEauthorrefmark{1},
        Julian Moosmann\;\orcidlink{0009-0007-0283-0031}\,\IEEEauthorrefmark{1},
        Andrea Gubler\,\IEEEauthorrefmark{2}\\
        Christian Vogt\;\orcidlink{0000-0003-4551-4876}\,\IEEEauthorrefmark{1},
        Florian J. Wegehaupt\;\orcidlink{0000-0002-3972-0561}\,\IEEEauthorrefmark{2},
        Michele Magno\;\orcidlink{0000-0003-0368-8923}\,\IEEEauthorrefmark{1}
        }
    \vspace{1mm}
    \IEEEauthorblockA{
        \IEEEauthorrefmark{1}\,\textit{Department of Information Technology and Electrical Engineering, ETH Zurich, Zurich, Switzerland}\\
        \IEEEauthorrefmark{2}\,\textit{Clinic of Conservative and Preventive Dentistry, Centre for Dental Medicine, University of Zurich, Zurich, Switzerland}
        }
}

\maketitle

\begin{abstract}
Dental caries is one of the most common chronic diseases worldwide, caused by acid production from bacterial metabolism of fermentable carbohydrates and affecting people of all ages. To evaluate the cariogenic and erosive properties of widely consumed food products, such as energy drinks, intraoral pH changes are measured during consumption.
The gold standard for such measurements is miniaturized silicon–lithium–barium glass membrane electrodes. These electrodes allow dental plaque to form on their surface, thereby enabling in situ monitoring of pH changes in a biologically relevant environment.
Due to their high impedance and susceptibility to external interference, they can currently only be measured using a large analog amplification and recording unit, which is highly limiting for study design and participant comfort, as individual measurements can take upwards of an hour.
This work presents the first battery-powered, low-power wireless and wearable pH telemetry evaluation system designed for real-time intraoral pH monitoring with glass electrodes.
The system comprises a miniaturized pH telemetry front-end, a neck-worn \ac{BLE} node, and software tools for data acquisition, visualization, and reporting.
The front end integrates with a custom dental prosthesis, directly digitizing the pH signal in the mouth and minimizing noise.
The data is transmitted over \ac{BLE} to a host computer, and analyzed using dedicated software that supports calibration, drift compensation, region marking, and PDF report generation.
The system integrates an 8.6 by 3.3\,mm, 0.2\,g pH front-end and a 37.6\,g neck-worn \ac{BLE} node which consume
8.89\,mW to transmit data at 10\,Hz to a host computer during a measurement.
A five-hour continuous validation using standard calibration solutions (pH 4, 7, 10) demonstrates system stability with a drift of only 0.005\,pH/min, and an effective response rate of 1.875\,pH/s, confirming robust sensing and reliable wireless telemetry. 
\end{abstract}

\vspace{5pt}

\begin{IEEEkeywords} 
Intraoral Telemetry,
Carcinogenicity Evaluation,
pH Sensing,
Wireless Data Acquisition,
Wearable 
\end{IEEEkeywords}             

\section{Introduction}\label{sec:intro}
Sugar, when combined with insufficient oral hygiene, is the primary contributor to dental caries and tooth decay in today's society~\cite{WHO2025_SugarsCaries}.
Almost immediately after food consumption, plaque bacteria begin producing intra-plaque acid by fermenting carbohydrates.
This acid first attacks the enamel and gradually demineralizes the dentin, the tooth's core, over time, resulting in caries.
Untreated, caries can lead to complications such as abscesses, toothaches, and tooth loss, and can also allow bacteria to enter the bloodstream, risking systemic infection.
Besides poor oral hygiene through insufficient brushing or rinsing, the consumption of sugary foods is considered to be the leading cause of the decay of dental health~\cite{Liu2025_MicrobiomeCaries}.
Thus, minimizing the consumption of fermentable sugars and carbohydrates is a direct approach to caries prevention~\cite{Large2024_ImpactUnhealthyFood}, enabled by proper quantification of the erosive properties of food products.

\begin{figure}
    \centering
    \begin{overpic}[width=1\columnwidth]{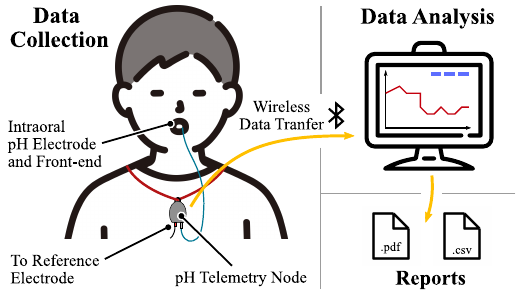}
        \put(0,53){(a)}
        \put(62,53){(b)}
        \put(62,16){(c)}
    \end{overpic}
    \vspace{-8mm}
    \caption{Conceptual overview of the proposed intraoral pH telemetry system. It consists of: (a) the data acquisition hardware, (b) the data analysis software, and (c) report generation. }
    \vspace{-7mm}
    \label{fig:ph_telemetry_overview}
\end{figure}

This is achieved using a method called \textit{pH-Telemetry}:
Plaque pH is measured in vivo during and for thirty minutes after consumption of the tested food product, using an indwelling intraoral pH electrode. By analyzing the pH curve obtained, the cariogenicity level of the product is determined~\cite{Toothfriendly2015}.
The gold standard for such measurements is glass electrodes with a thin silicon-lithium-barium membrane (green tip), which is placed in the participant's mouth~\cite{intra_oral_wire_telemetry_1977}.
A second reference electrode is placed on the participant's arm, which features four porous membranes that enable the exchange of ions, closing the circuit formed between the two electrodes through the electrically conductive human body and enabling oral pH sensing~\cite{identification_caries_risks_1983, intra_oral_wire_telemetry_1977}.
In addition to their accuracy, glass electrodes are uniquely capable of measuring the effects of plaque formation caused by insufficient oral hygiene. By embedding them into custom dental prosthetics, participants can wear them continuously for long periods of time before the measurement, allowing plaque to form directly on the electrode.
This direct pH measurement approach in the plague's proximity minimizes the measurement difference between plague formation on teeth and plague formation on the sensing element, allowing for representative in-situ results.\cite{glass_electrode_pandolfino_2006, LI2022100135, intraoral_ph_abelson_1981,identification_caries_risks_1983, intra_oral_wire_telemetry_1977}.

Glass electrodes, however, are highly susceptible to noise due to their significant output impedance (typically in the range of \qty{}{\giga\ohm}~\cite{skoog2019textbook}) and low output amplitude ~\cite{pH_electrode_comp_smit_1997}, requiring a very stable, low-noise amplifier front-end. 
Currently, the only available unit capable of interfacing with such electrodes is a large desktop apparatus, feeding an analog chart recorder.
Because individual measurements often last an hour or more, and experiments may require repeated measurements over the course of weeks, the cumbersome measurement process can be uncomfortable for participants, and limits study design. 
Furthermore, the analog data handling and report preparation involve several steps, making their generation a demanding process.
These limitations highlight the need for a fully digital, compact, and user-friendly intraoral pH telemetry system capable of providing real-time monitoring and automated data handling. 
Although several digitalization approaches for intraoral pH monitoring have been presented~\cite{timpel2023sensors, tooth_monitoring_tseng_2018}, none of them are compatible with silicon-lithium-barium membrane glass electrodes. 

This work proposes a fully digital intraoral pH telemetry platform that combines a high-impedance frontend with direct intraoral data digitization to optimize signal quality and minimize noise. Once digitized, the data is wirelessly transmitted to a computer, improving user comfort while enabling unobtrusive, real-time measurement control and observation \cite{oral_ph_telemetry_ro_2007} with best-in-class glass electrodes. 
In particular, this article presents the following contributions:
\begin{enumerate}
    \item \textbf{Miniaturized high-impedance front-end:} A custom front-end for a miniaturized glass electrode that combines a high-impedance frontend with direct intraoral data digitalization to optimize signal quality and fast settling times for continuous intraoral pH monitoring.
    \item \textbf{Wireless telemetry:} A wireless node enables real-time data forwarding for intraoral pH telemetry.
    \item \textbf{Digital end-to-end pipeline:} A full end-to-end system for continuous intraoral pH monitoring, including data analysis and report generation.
\end{enumerate}

\section{System Design}\label{sec:methods}
The wireless and wearable pH telemetry system comprises three functional units: the pH telemetry frontend \cref{subsec:frontend}, the pH telemetry node \cref{subsec:node}, and the reporting software \cref{subsec:software}.
Together, they provide a complete end-to-end data analysis and processing chain for measuring, forwarding, analyzing, and report generation, as shown in \cref{fig:ph_telemetry_overview}.

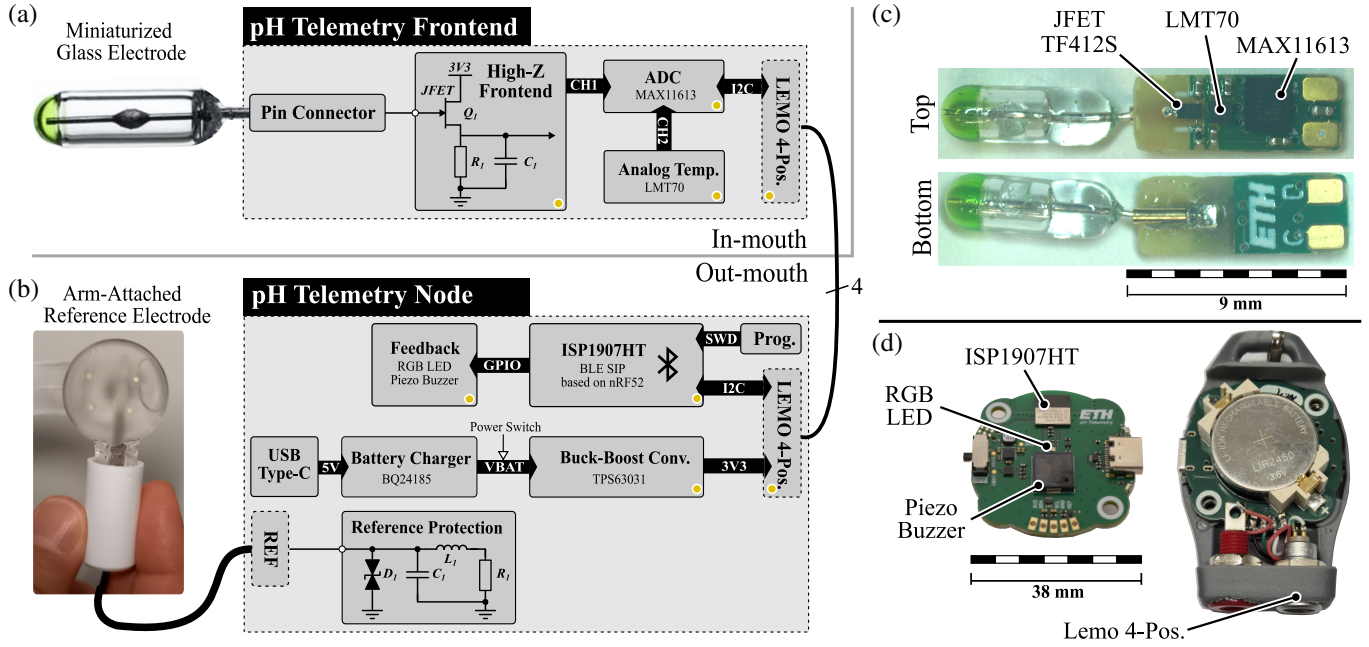
\begin{figure*}[!t]
    \centering
    \begin{overpic}[width=\textwidth]{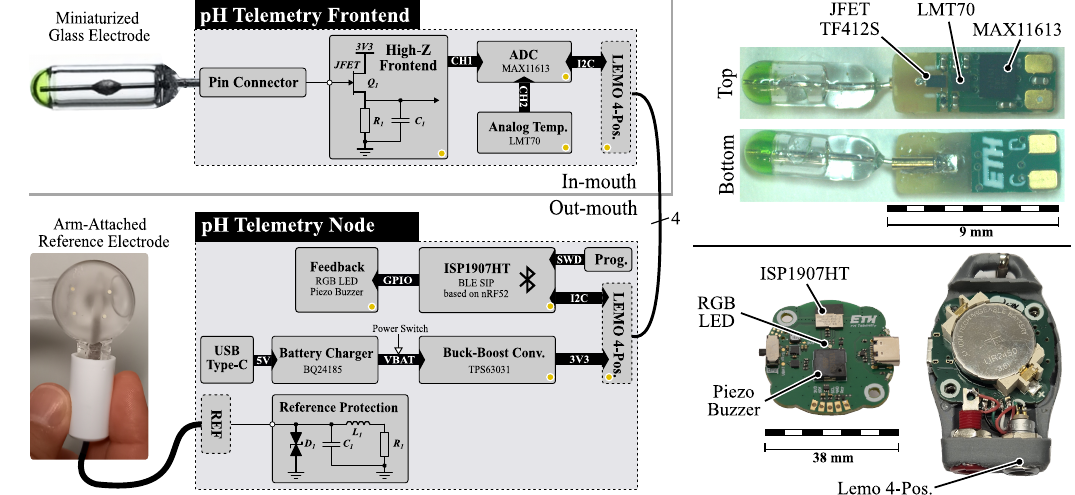}
        \put(1,45){(a)}
        \put(1,25){(b)}
        \put(64,45){(c)}
        \put(64,21){(d)}
    \end{overpic}
    \vspace{-6mm}
    \caption{System overview: (a) High-level block diagram of the high impedance pH telemetry front-end located in the mouth, 
    (b) High-level block diagram of the pH telemetry node for data forwarding, (c) Top- and bottom-view of the implemented telemetry front-end, (d) Implementation of the pH telemetry node.}
    \vspace{-6mm}
    \label{fig:ph_telemetry_electronic}
\end{figure*}

\subsection{pH Telemetry Front End}\label{subsec:frontend}
\subsubsection{Electrical Design}
At the heart of the system sits a miniaturized glass electrode manufactured by Willi Möller AG, located in Oerlikon, Switzerland \cite{willi_moeller_ag}, and shown in \cref{fig:ph_telemetry_electronic} (a) \& (c).
The pH telemetry front-end is responsible for sensing, conditioning, and digitizing the electrode's analog pH potential.
The primary design challenge arises from the electrode's extremely high source impedance of \qty{5.1}{\giga\ohm}, which makes the analog pH potential very susceptible to noise and loading by external circuitry.
To minimize external interference, the entire front-end must be miniaturized, enabling it to be positioned as close as possible to the electrode.
This is achieved using a high-impedance input buffer built on a \ac{JFET} in source follower (common drain) configuration, see \cref{fig:ph_telemetry_electronic} (a).
The frequency-dependent input impedance of this buffer is dominated by capacitive coupling, and is given by \cref{eq:input_impedance}:
\begin{equation}
\label{eq:input_impedance}
    C_{\mathrm{iss}} = C_{gs}+C_{gd} \qquad
    |Z_{\mathrm{in}}(f)| = \frac{1}{2\pi f\,C_{\mathrm{iss}}}
\end{equation}
Here, $C_{gs}$ and $C_{gd}$ are the gate-source and gate-drain capacitances, with $C_{\mathrm{iss}}$ being the total \ac{JFET} input capacitance.
The specific \textit{On Semiconductor TF412S} \ac{JFET} selected, presents a \(C_{\mathrm{iss}}\) of  \(4\;pF\), giving the amplified an input impedance of up to \qty{40}{\giga\ohm} for frequencies below \qty{1}{\hertz}. This is sufficiently high to ensure minimal electrode loading and thus minimize its effect on the pH-sensing process. To facilitate electrical connection and eliminate the need for soldering, the miniaturized glass electrode is interfaced with the front-end via a \textit{MILL-MAX} connector.
The buffered analog signal is then passed through an RC low-pass filter before being routed to the \textit{Analog Devices MAX11613} 12-bit \ac{ADC} for digitization.
Because signal conditioning and digitization are performed directly in the front-end intraorally, environmental noise is minimized. 
Next to the pH sensor signal, an analog temperature sensor is connected to the same \ac{ADC}, enabling simultaneous temperature measurements with the pH signal. This allows compensation for temperature-dependent variations in pH readings. 
The digitized outputs of both pH and temperature are made accessible to the pH telemetry node via the \ac{ADC}'s \ac{I2C} bus.

\subsubsection{Hardware Implementation}
The hardware realization of the pH telemetry front end, depicted in \autoref{fig:ph_telemetry_electronic} (c), is implemented on a 2-layer, \qty{0.8}{\milli\meter} thick FR-4 \ac{PCB}, measuring \qty{8.6}{\milli\meter} by \qty{3.3}{\milli\meter}, with a total weight of only \qty{0.2}{\gram}. 
It incorporates a spring-loaded connector to electrically interface with the miniaturized pH glass electrode, preventing \ac{KCl} evaporation during soldering. To minimize leakage currents and preserve signal integrity, the copper layer and solder mask surrounding the \ac{JFET} buffer have been selectively removed, thereby increasing the local surface resistivity.
The underlying FR-4 substrate offers a surface sheet resistivity \((\rho S)\) of approximately \qtyrange{10}{100}{\giga\ohm\persquare} and a volume resistivity \((\rho V)\) in the order of \qty{100}{\giga\ohm/\centi\meter}~\cite{ipc4101c}, which is sufficiently large compared to the electrode's source impedance of \qty{5.1}{\giga\ohm} to ensure low-leakage operation for this application.
Electrical connectivity is provided via four gold-plated contact pads, distributed symmetrically across both sides of the printed circuit board. These expose power (\qty{3.3}{\volt}, marked \textit{+}), and ground (marked \textit{-}) on the top side, and the \ac{I2C} clock and data lines (labeled textit{C} and \textit{D} respectively), on the bottom side. All surface-mounted components are populated on the top side of the PCB, simplifying assembly and manufacturing.
\subsection{pH Telemetry Node}\label{subsec:node}
\subsubsection{Electrical Design}
The pH telemetry node serves as the central unit, controlling and powering the wearable wireless in-mouth pH telemetry system.
It interfaces with the pH telemetry front-end over \ac{I2C} and wirelessly forwards the data via \ac{BLE} to a host computer.
\autoref{fig:ph_telemetry_electronic} (b) shows a simplified block diagram of the pH telemetry node.
At the heart of the system is the \textit{Insight SIP ISP1907HT} \ac{BLE} \ac{SiP}, which integrates a \textit{Nordic Semiconductor nRF52833} ARM Cortex-M4 \ac{MCU}, along with RF matching circuitry, an integrated antenna, and both \qty{32}{\mega\hertz} and \qty{32}{\kilo\hertz} crystal oscillators for precise timing and communication.
The node connects to the pH telemetry frontend via a 4-pin \textit{LEMO} connector, which carries the power lines (VCC, GND) and \ac{I2C} communication signals (SDA, SCL).
A rechargeable and replaceable Li-Ion battery powers the wireless system. Batter charging and protection is handled by the \textit{Texas Instruments BQ25185} battery management IC. A hardware power switch enables manual activation and shutdown of the system.
When turned on, a \textit{Texas Instruments TPS63031} buck-boost converter regulates the battery voltage to a stable \qty{3.3}{\volt}, ensuring consistent system performance independent of the battery’s state of charge. An integrated RGB LED and buzzer module provides visual and audible feedback during operation, enhancing user interaction and status indication.
To accommodate the required arm-attached reference electrode, a dedicated reference connector (REF) is routed to a 4 mm banana jack, which is internally protected against electrostatic discharge (ESD) and DC overcurrent. 

\subsubsection{Hardware Implementation}
The hardware implementation of the pH telemetry node is shown in \autoref{fig:ph_telemetry_electronic}(d).
The electronics are encapsulated in a compact, necklace-shaped enclosure.
At the bottom of the enclosure, two connectors, the \qty{4}{\milli\meter} banana jack and the 4-pin  \textit{LEMO} connector, are mechanically fixed to the housing and electrically interfaced to the main \ac{PCB} via flexible strand wires. 
Their mass, particularly when external cables or electrodes are attached, helps stabilize the device's orientation during use, ensuring consistent positioning on the chest while wearing.
The housing's sidewalls include a slide-on/off switch and a USB Type-C connector for charging. Adjacent to the charging port, two status LEDs (red and green) indicate the charging state and signal potential fault conditions.
An opening in the center of the enclosure provides direct visibility of the onboard RGB LED, which conveys system status and feedback during operation.The complete pH telemetry node measures approximately \qty{40}{\milli\meter} in width, \qty{22}{\milli\meter} in height, and \qty{68}{\milli\meter} in depth, with a total weight of \qty{37.61}{\gram}, including the battery and connectors.

\subsubsection{Firmware}
The firmware running on the pH telemetry node is based on the Zephyr \ac{RTOS} and implements the core functionalities for pH frontend interfacing, Bluetooth communication, and user feedback. Analog data from the high-impedance pH glass electrode and the temperature sensor are sampled at \qty{100}{\hertz} using the 12-bit \textit{Analog Devices MAX11613} \ac{ADC}. 
The acquired pH and temperature readings are averaged over 10 samples and then filtered with a moving average to reduce noise.
The processed data are then wirelessly transmitted to a host computer via \ac{BLE} at an update rate of \qty{10}{\hertz} (every \qty{100}{\milli\second}). This sampling rate is sufficient to observe the measurements in real time and to capture rapid changes, e.g., caused by sugary beverages \cite{sugary_beverages_salvia_rinki_2016} that are typically below \qty{1}{\pH/\min} \cite{in_vivo_pH_sarah_2025}.

\subsection{Reporting Software}\label{subsec:software}
Furthermore, the custom recording and analysis software developed simplifies the entire workflow. 
The first connects to the device via \ac{BLE}, and records the measurement. The data is visualized in real time, and the operator is able to annotate different experimental phases for later post-processing.
The second software allows for data inspection, automatic drift compensation, post-processing, and report generation.

\section{Experimental Setup \& Results}\label{sec:experimental_setup}
To assess the performance and reliability of the pH telemetry system, a series of tests was conducted under controlled laboratory conditions.
The evaluation focused on verifying the stability and responsiveness of the system using standardized pH calibration solutions at pH 4, pH 7, and pH 10.

\subsection{Test Setup}
Instead of the arm-attached electrode, an Ag/AgCl reference electrode of the same construction, but with a single membrane instead of four filled with 3 M \ac{KCl} solution, is used to evaluate the system.
It is connected via a \qty{1}{\meter} flexible stranded wire terminated with a \qty{4}{\milli\meter} banana plug, as used by the arm-attached reference electrode.
The pH telemetry frontend is connected to the pH telemetry node via a \qty{1.6}{\meter} long four-core cable.
The individual wires are directly soldered to the exposed signal pins of the frontend.

\begin{figure}[h]
    \centering
    \vspace{-2mm}
    \includegraphics[width = 1\columnwidth]{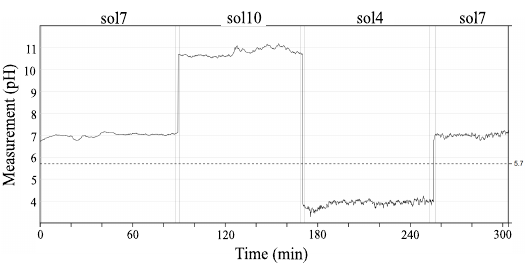}
    \vspace{-8mm}
    \caption{Results of the 5-hour evaluation using the pH sequence \(7\rightarrow10\rightarrow4\rightarrow7\) generated by the reporting software.}
    \label{fig:test_results}
\end{figure}

Before each measurement, the miniaturized pH glass electrode was inspected and filled with 3 M \ac{KCl} under vacuum to ensure stable ionic conductivity and reference potential. It was then soaked in distilled water for 1 h to hydrate the glass membrane and wet the reference junction.
This preparation is critical for restoring the hydrated layer required for accurate, responsive pH measurements. For data acquisition, the pH telemetry node was powered on, and both electrodes were immersed in pH buffer solutions dispensed into petri dishes.

Measurements were recorded over 5 hours to validate the correct functioning of the sensing frontend, data transmission via \ac{BLE}, and real-time visualization and logging through the reporting software; see \cref{fig:test_results}.
The measurement started by creating the first calibration point by dipping the electrodes into the pH 7 calibration solution. Next, the solutions were adjusted to pH 10 and then to pH 4 to assess the system's responsiveness. A second exposure to pH 7 provides a second reference point for calibrating the linear pH-potential drift, which is a common phenomenon of pH glass electrodes \cite{pH_electrode_alnoush_2021}.
To compensate for this drift, the measured electrode potential was linearly corrected between the two pH 7 reference points, effectively applying a two-point time-based calibration by averaging the pH-potential values from each calibration phase.

Power consumption was measured using the \textit{Nordic Semiconductor nRF Power Profiler Kit II} at a constant supply voltage of \qty{3.3}{\volt}. Total system power was recorded in source-meter mode, while individual component currents were measured in ampere-meter mode.

\subsection{Results}\label{sec:results}
The conducted experiments show that the measurement remains stable in a pH 7 solution, with the measured potential varying by less than \qty{\pm0.15}{\pH} over a measurement duration of \qty{90}{\minute}.
The investigation of the slope steepness of the transition from pH 10 to pH 4 shows that the pH potential stabilizes after \qty{3.2}{\second}, resulting in an effective response rate of \qty{1.875}{\pH/\second}.
Comparing the average values from the initial and ending reference point after 5 hours of constant measurement revealed a linear drift of \qty{0.005}{\pH/\minute} and a pH electrode sensitivity of only \qty{31}{\milli\volt/\pH}. The reduced sensitivity below the theoretical Nernst slope of \qty{59.16}{\milli\volt/\pH} at room temperature (\qty{25}{\celsius}) is most likely caused by electrode aging.
Based on these results, it can be concluded that the measurement electronics function as expected. The pH potential remains stable over extended periods. Furthermore, the complete pH telemetry measurement chain is verified, comprising the pH front-end, \ac{BLE} node, and the recording and analysis software.

The power analysis yields the current consumption values summarized in~\cref{tab:power_consumption}.
The total system power consumption at \qty{3.3}{\volt} amounts to \qty{15.81}{\milli\watt}. By deactivating the status led, the system's power consumption can be reduced to \qty{8.89}{\milli\watt}.
\begin{table}[h]
    \centering
    \renewcommand{\arraystretch}{1.3}
    \vspace{-6mm}
    \caption{Power consumption of active components at \qty{3.3}{\volt}}
    \label{tab:power_consumption}
    \setlength{\tabcolsep}{5pt}
    \renewcommand{\arraystretch}{1.2}
    \begin{tabularx}{\columnwidth}{
    >{\hsize=10\hsize\centering\arraybackslash}X
    >{\hsize=11\hsize\centering\arraybackslash}X
    >{\hsize=7\hsize\centering\arraybackslash}X
        }
        \toprule
        \multirow{2}{*}{\textbf{Component}} &
        \multirow{2}{*}{\textbf{Part Name}} &
        \textbf{Power Consumption} \\
        \midrule
        ADC (intraoral)                     & MAX11613      & \qty{1.09}{\milli\watt} \\    
        Temp. Sensor (intraoral)            & LMT70         & \qty{49}{\micro\watt}  \\     
        Front-end (intraoral)               & TF412         & \qty{396}{\micro\watt}  \\    
        Microcontroller                     & ISP1907HT     & \qty{7.35}{\milli\watt} \\    
        Status LED (blue)                   & SML-LX0404SIUPGUSB & \qty{6.93}{\milli\watt} \\    
        \midrule
        \textbf{Full system} & & \textbf{\qty{15.81}{\milli\watt}} \\
        \bottomrule
    \end{tabularx}
    \vspace{-5mm}
\end{table}

\section{Conclusion}\label{sec:conclusion}
This paper presents the design, implementation, and evaluation of an end-to-end system for wireless intraoral pH Telemetry. The system integrates an \qty{8.6}{\milli\meter} by \qty{3.3}{\milli\meter}, \qty{0.2}{\gram} pH front-end and a \qty{37.6}{\gram} neck-worn \ac{BLE} node which consume \qty{8.89}{\milli\watt} to transmit data at \qty{10}{\hertz} to a host computer for real-time measurement observation and data annotation.
Thanks to the combination of a high-impedance front end and an ADC both located in the mouth, signal quality improves, resulting in a pH potential variation of less than \qty{\pm0.15}{\pH} over \qty{90}{\minute}. A five-hour long-term validation under laboratory conditions using demonstrated system stability, revealing a drift of only \qty{0.005}{\pH/\minute} and an effective response rate of \qty{1.875}{\pH/\second}.

\begin{acronym}

    \acro{BAN}{Body Area Network}
    \acro{WBAN}{Wireless Body Area Network}
    \acro{PAN}{Personal Area Network}
    \acro{IoT}{Internet of Things}
    \acro{IoB}{Internet of Bodies}
    \acro{AI}{Artificial Intelligence}
    \acro{MCU}{Microcontroller}
    \acro{GNSS}{Global Navigation Satellite System}
    \acro{Nb-IoT}{Narrowband IoT}
    \acro{LoRa}{Long Range}
    \acro{LoRaWAN}{Long Range Wide Area Network}
    \acro{BLE}{Bluetooth Low Energy}

    \acro{NFC}{near Field Communication}
    \acro{SpO2}{Oxigen Saturation}
    \acro{VR}{Virtual Reality}
    \acro{AR}{Augmented Reality}
    \acro{EQS-HBC}{Electro-Quasistatic Human Body Communication}
    \acro{EQS}{Electro-Quasistatic}
    \acro{HBC}{Human Body Communication}
    \acro{WBAN}{Wireless Body Area Network}
    \acro{WPAN}{Wireless Personal Area Network}
    \acro{WP}{Work Package}
    \acro{CSMA}{Carrier Sense Multiple Access}
    \acro{KB}{Kilobyte}
    \acro{PMIC}{Power Management IC}

    \acro{HCI}{Human-Computer Interaction}
    \acro{HMI}{Human-Machine Interaction}
    \acro{TIA}{Transimpedance Amplifier}
    \acro{SoTA}{State-of-The-Art}
    \acro{SoA}{State-of-Art}
    \acro{WLAN}{Wireless Local Area Network}
    \acro{PCE}{power Conversion Efficiency}
    \acro{ECG}{Electrocardiogram}
    \acro{BOM}{Bill of Material}
    \acro{DAQ}{Data Acquisition}
    \acro{RSSI}{Received Signal Strength Indicator}
    \acro{RSS}{Received Signal Strength}
    \acro{GBP}{Gaind-Bandwidth Product}

    \acro{LoS}{Line-of-Sight}
    \acro{nLoS}{non-Line-of-Sight}
    \acro{QoS}{Quality of Service}
    \acro{NB}{Narrowband Communication}
    \acro{WPT}{Wireless Power Transfer}
    \acro{IBPT}{Intra-Body Power Transfer}
    \acro{ICNIRP}{International Commission on Non-Ionizing Radiation Protection} 
    \acro{SAR}{Specific Energy Absorption}

    \acro{RF}{Radio Frequency}
    \acro{RFID}{Radio Frequency Identification}
    \acro{IoT}{Internet of Things}
    \acro{IoUT}{Internet of Underwater Things}
    \acro{UWN}{Underwater Wireless Network}
    \acro{UWSN}{Underwater Wireless Sensor Node}
    \acro{AUV}{Autonomous Underwater Vehicles}
    \acro{UAC}{Underwater Acoustic Channel}

    \acro{ASK}{Amplitude Shift Keying}
    \acro{UUID}{Universal Unique Identifier}
    \acro{PZT}{Lead Zirconium Titanate}
    \acro{AC}{Alternating Current}
    \acro{NVC}{Negative Voltage Converter}
    \acro{NVCR}{Negative Voltage Converter Rectifier}
    \acro{FWR}{Full-Wave Rectifier}
    \acro{GPIO}{General Purpose Input/Output}
    \acro{PCB}{Printed Circuit Board}
    \acro{AUV}{Autonomous Underwater Vehicle}
    \acro{IMU}{Intertial Measurement Unit}
    \acro{BLE}{Bluetooth Low Energy}
    \acro{FSR}{Force Sensing Resistor}
    \acro{SiP}{System in Package}
    \acro{SoC}{System on Chip}
    \acro{SpO2}{Oxigen Saturation}
    \acro{PULP}{Parallel Ultra-Low Power}
    \acro{ML}{Machine Learning}
    \acro{ADC}{Analog to Digital Converter}
    \acro{TCDM}{Tightly Coupled Data Memory}
    \acro{GNSS}{Global Navigation Satellite System}
    \acro{IC}{Integrated Circuit}
    \acro{POM}{Polyoxymethylene}
    \acro{RTOS}{Real-Time Operating System}
    \acro{LoRaWAN}{Long Range Wide Area Network}
    \acro{ML}{Machine Learning}
    \acro{IIR}{Infinite Impulse Response}
    \acro{PLL}{Phase-locked Loop}

    \acro{KCl}{potassium chloride}

    \acro{I2C}{Inter-Integrated Circuit}
    \acro{JFET}{Junction Field-effect Transistor}

    \acro{MPPT}{Maximum Power Point Tracking}
    \acro{MPP}{Maximum Power Point}

    \acro{MMN}{Medical Micropower Network}
    \acro{MBAN}{Medical Body Area Network}
    \acro{WMTS}{Wireless Medical Telemetry Service}
    \acro{UWB}{Ultra-Wideband}
    \acro{BLE}{Bluetooth Low Energy}

    \acro{FSK}{Frequency Shift Keying}
    \acro{OOK}{On-Off Keying}
    \acro{PSK}{Phase Shift Keying}
    \acro{QAM}{Quadrature Amplitude Modulation}
    \acro{BER}{Bit Error Rate}
    \acro{SNR}{Signal-to-Noise Ratio}
    \acro{LTC}{Linear Timecode}

    \acro{SNSF}{Swiss National Science Foundation}

    \acro{OTS}{off-the-shelf}
    
\end{acronym}

\section*{ACKNOWLEDGMENT}
This work was partially funded by the CHIST-ERA project ”SNOW” (Grant 209675).
\newpage
\bibliographystyle{IEEEtranDOI} 

\bibliography{  
    bib/iscas
}

\end{document}